\documentclass[12pt]{article}
\usepackage{a4wide,epsfig,amsmath,amssymb,cite}
\usepackage{scalefnt}
\newcommand{\Lmu}{L_\mu}

\newcommand{\order}[1]{\mathcal{O}\left( #1 \right)}

\newcommand{\VerttopOSlnpb}{V_{np2b2}}
\newcommand{\VerttopOSlnp}{V_{np12}}
\newcommand{\VerttopOSlnpnp}{V_{np33}}

\begin{document}


\title{\vskip-3cm{\baselineskip14pt
\begin{flushleft}
\normalsize Alberta Thy 06-09 \\
\normalsize SFB/CPP-09-30 \\
\normalsize TTP09-09 \\
\end{flushleft}}
\vskip1.5cm
Completely automated computation of the heavy-fermion corrections to the
three-loop matching coefficient of the vector current
}

\author{\small 
  P. Marquard$^{(a)}$, J.H. Piclum$^{(b)}$, D. Seidel$^{(b)}$,
  M. Steinhauser$^{(a)}$
\\
{\small\it (a) Institut f{\"u}r Theoretische Teilchenphysik,
  Universit{\"a}t Karlsruhe (TH)}\\ 
{\small\it Karlsruhe Institute of Technology (KIT), 76128 Karlsruhe, Germany}
\\
{\small\it (b) Department of Physics, University of Alberta, Edmonton,
  Alberta, Canada T6G 2G7}
}

\date{}

\maketitle

\thispagestyle{empty}

\begin{abstract}
We evaluate the corrections to the matching coefficient of the vector
current between Quantum Chromodynamics (QCD) and Non-Relativistic QCD
(NRQCD) to three-loop order containing a closed heavy-fermion loop.
The result constitutes a building block both for the bottom- and top-quark
system at threshold. Strong emphasis is put on our completely automated
approach of the calculation including the generation of the
Feynman diagrams, the identification of the topologies, the reduction to
master integrals and the automated numerical computation of the latter.

\medskip

\noindent
PACS numbers: 12.38.Bx 12.38.-t 14.65.Ha
\end{abstract}

\newpage


\section{Introduction}

A major goal of a future international linear collider (ILC) is
the precise measurement of the top-quark production cross section
close to threshold. 
Next to a precise extraction of the strong coupling, an unrivalled
determination of the top-quark mass and its width is possible.
This would open up a new chapter in the electroweak precision physics
which leads to very strong checks of the Standard Model or
possible extensions.

The theoretical calculation of the threshold cross section is 
based on an effective theory~\cite{Caswell:1985ui,Bodwin:1994jh} (for
a review see~\cite{Brambilla:2004wf}) which is constructed from 
QCD by integrating out the hard scale given by the top-quark mass.
The connection between the two theories is established by so-called
matching coefficients which constitute the coupling constants of the effective
operators within NRQCD.

A preliminary analysis to next-to-next-to-next-to leading order
(NNNLO) of the top-quark threshold cross section, which is necessary in order
to match the expected experimental precision~\cite{Martinez:2002st}, has been
performed in Ref.~\cite{Beneke:2008ec}. However, the three-loop static
potential (see Ref.~\cite{Smirnov:2008pn} for the fermionic contribution) and
the three-loop matching coefficient beyond the light-fermion
approximation~\cite{Marquard:2006qi} are still missing. In this paper we
provide a further building block needed for the completion of the NNNLO
calculation: the heavy-fermion contribution to the three-loop matching
coefficient.
Next to important applications in the top-quark sector NRQCD is also
an appropriate tool for the description of boundstate phenomena of 
charm and bottom quarks~\cite{Brambilla:2004wf}. 

The matching coefficient of the vector current 
in the full and effective theory is defined through ($k=1,2,3$)
\begin{eqnarray}
  j^k_v &=& c_v(\mu) \tilde{j}^k 
  + {\cal O}\left(\frac{1}{m_Q^2}\right)
  \label{eq::def_of_cv}
  \,,
\end{eqnarray}
where the vector current in the full and effective theory reads
\begin{eqnarray}
  j_v^\mu = \bar{Q} \gamma^\mu Q\,,
  \qquad
  \tilde{j}^i = \phi^\dagger \sigma^i \chi\,.
\end{eqnarray}
$Q$ denotes a generic heavy quark with mass $m_Q$ and
$\phi$ and $\chi$ are two-component Pauli spinors for quark and
anti-quark, respectively.
In this paper we compute the three-loop non-singlet contribution to $c_v$
which contains one or two closed heavy quark loops.
The analog corrections involving closed light (massless) quark loops
have been considered in Ref.~\cite{Marquard:2006qi}.

The evaluation of the three-loop diagrams contributing to $c_v$ is
quite involved. One has to consider vertex diagrams with massive
quarks on their mass shell and the external momentum $q^2=4m_Q^2$. 
This kinematical configuration in
combination with an involved reduction to master integrals makes 
the calculation quite challenging.
In this paper we discuss an automated setup which minimizes
the manual interaction. Even the results for the master
integrals are obtained in an automated way.

Our approach for the automated calculation is introduced in the next
section. Afterwards we present the results for the matching coefficient
in Section~\ref{sec::match} and conclude in Section~\ref{sec::concl}.


\section{Automated multi-loop calculation}

When evaluating multi-loop Feynman integrals one encounters several
difficulties which have to be overcome. Among them are the generation of the
Feynman diagrams, the reduction of the many integrals which appear at the
initial stage of the calculation to a relatively small set of so-called master
integrals and the evaluation of the latter.
Very often the individual steps are automated, however, the interplay between
them is not. In the following we present a
setup where various program packages are combined in order to minimize the
manual work.

The individual steps of our automated setup are as follows
\begin{enumerate}
\item All Feynman diagrams are generated with {\tt
    QGRAF}~\cite{Nogueira:1991ex} which requires two input files: one
  specifying the process and one containing the propagators and vertices
  occurring in the theory.
\item The output of {\tt QGRAF} is transformed to {\tt FORM}~\cite{Vermaseren:2000nd} with
  the help of the program {\tt q2e}~\cite{Harlander:1997zb,Seidensticker:1999bb}.
  {\tt q2e} requires as input the {\tt FORM} notation for the propagators and
  vertices and information about the hierarchy of the particle masses 
  and the external momenta.
\item The output of {\tt q2e} is further processed with {\tt exp} which 
  identifies the various topologies and generates for each diagram a 
  separate file containing complete information like the projectors to be
  applied, the expansion to be performed and the topology file to be called.
\item In a next step the {\tt FORM} part is initiated. After taking the
  traces and applying the projectors a topology-specific file is included
  which expresses the result in terms of a sum of scalar integrals.
\item From the sum of all diagrams a list of integrals is extracted. This list
  serves as input for {\tt crusher}~\cite{PMDS} 
  which produces for each topology a table containing the
  reduction to master integrals.
\item 
  Each topology produces a certain number of master integrals. A small
  {\tt Mathematica} routine combines all master integrals, identifies
  identical ones and generates relations among them.
\item The tables from step~5
  together with the relations among master integrals (step~6)
  are applied to the sum of the bare diagrams leading to a
  representation of the result in terms of a minimal set of master integrals 
  multiplied with $\epsilon$-dependent coefficients.
\item{}
  The input for {\tt crusher} can also be used 
  for {\tt FIESTA}~\cite{Smirnov:2008py} which we employ
  in order to obtain numerical results for the master integrals.
\item At four places information about the topologies are needed: as input
  for {\tt exp} (cf. step~3) and {\tt crusher}
  (cf. step~5), for the topology-specific
  {\tt FORM} file (cf. step~4) and for identifying identical master integrals
  (cf. step~6). 
  This input is generated automatically from a file
  containing the
  definition of the three-loop topologies for the two-point on-shell integrals
  entering, e.g., the $\overline{\rm MS}$-on-shell relation of the quark
  mass~\cite{Chetyrkin:1999qi,Chetyrkin:1999ys,Melnikov:2000qh,Marquard:2007uj}. 
  This file is available from our previous calculation~\cite{Marquard:2007uj}.
  Note that this is the only input which contains non-trivial problem-specific
  information; the remaining input files (see steps~1 and~2)
  are either quite generic (e.g. 
  identical for all QCD processes) or quite simple to adapt like the 
  specification of the process under consideration for {\tt QGRAF}.
\end{enumerate}

Let us stress that the automated setup outlined above requires little
interaction from outside and thus minimizes possible errors.

Many steps of the above list have been used extensively in previous
calculations, however, the automatic calculation of the master integrals is
new. Such an approach is essential 
in those cases where many master integrals occur.
In our calculation we encounter 24 master integrals 
for the contribution involving a closed heavy fermion loop.

Of course, a setup as described above requires several checks to be performed
on the final result. On one hand they certainly include gauge parameter
independence and the finiteness, on the other hand 
several checks on the numerical stability of our result are necessary.
In our case the latter include the following:
\begin{itemize}
\item
  {\tt FIESTA}~\cite{Smirnov:2008py}, which is an efficient implementation of
  the sector decomposition method for the evaluation of master integrals,
  provides the 
  possibility to introduce a lower cut-off in the numerical integration where
  numerical instabilities can occur. 
  Actually, for the evaluation of our master integrals
  we have to choose a non-zero cut-off. Its variation provides an 
  estimate of the uncertainty.
\item 
  {\tt FIESTA} furthermore provides an uncertainty from the underlying
  Monte-Carlo integration performed with {\tt Vegas}~\cite{Lepage:1977sw}
  which we also  take into account.
\item 
  A further estimate of the uncertainty is provided by changing the basis 
  used for the master integrals. The corresponding relations among the master
  integrals can be obtained in a straightforward way from the reduction tables
  generated by {\tt crusher}.
\item 
  Some of the master integrals
  are known analytically and can thus be used to replace the
  corresponding numerical expressions.
\end{itemize}


\section{\label{sec::match}Matching coefficient}

\begin{figure}
  \leavevmode
  \epsfxsize=\textwidth
 \epsffile{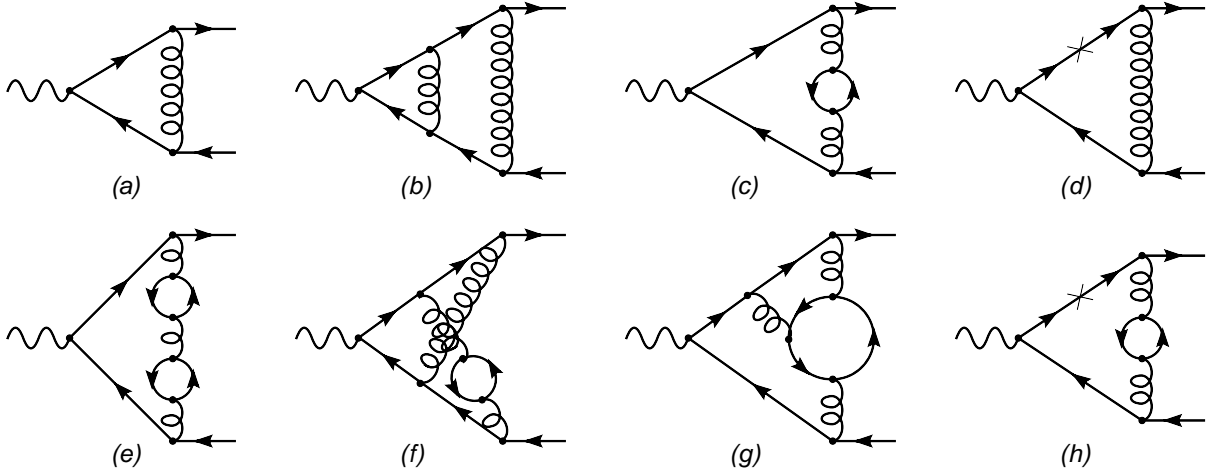}
\caption{\label{fig::sample} Feynman diagrams contributing to the 
  matching coefficient. Solid lines denote (heavy) quarks with mass $m_Q$
  and curly lines denote gluons. In (d) and (h) mass counterterm diagrams are
  shown.} 
\end{figure}

Starting from Eq.~(\ref{eq::def_of_cv}) it is possible to derive the
equation\footnote{See, e.g., 
  Ref.~\cite{Marquard:2006qi} for a derivation based on the threshold
  expansion~\cite{Beneke:1997zp,Smirnov:2002pj}.}  
\begin{eqnarray}
  Z_2 \Gamma_v &=& c_v \tilde{Z}_v^{-1}
  \,,
  \label{eq::def_cv}
\end{eqnarray}
where $\Gamma_v$ denotes the one-particle irreducible vertex diagrams
with on-shell quarks with momenta $q_1$ and $q_2$ and $q^2=(q_1+q_2)^2=
4 m_Q^2$. 
Some sample Feynman diagrams contributing at the one-, two- and three-loop 
level are shown in Fig.~\ref{fig::sample}. 
$Z_2$ is the wave function renormalization constant in the on-shell scheme and 
$\tilde{Z}_v$ collects the infra-red divergences within the minimal subtraction
scheme. The latter are cancelled against ultra-violet
divergences of the effective theory rendering physical quantities
finite.

The two- and three-loop corrections to $Z_2$ have been computed in
Refs.~\cite{Broadhurst:1991fy} 
and~\cite{Melnikov:2000zc,Marquard:2007uj}, respectively, and $\tilde{Z}_v$ can
be obtained from Ref.~\cite{Kniehl:2002yv}
(see also Refs.~\cite{Marquard:2006qi,Beneke:2007pj}).
The latter reads
\begin{eqnarray}
  \tilde{Z}_v &=& 1 + \left(\frac{\alpha_s^{(n_l)}(\mu)}{\pi}\right)^2 \left(
    \frac{1}{12} C_F^2 + \frac{1}{8} C_FC_A \right) \frac{\pi^2}{\epsilon}
  \nonumber \\
  &&\mbox{}
  + \left(\frac{\alpha_s^{(n_l)}(\mu)}{\pi}\right)^3 C_F T \Bigg\{
  n_l \left[ \left(
      \frac{1}{54} C_F + \frac{1}{36} C_A \right) \frac{\pi^2}{\epsilon^2} 
    -  \left( \frac{25}{324} C_F + \frac{37}{432} C_A \right)
    \frac{\pi^2}{\epsilon} \right] \nonumber \\
  &&\mbox{}
  + 
     C_F n_h \frac{\pi^2}{60\epsilon} 
  \Bigg\}
  + \dots \,, 
  \label{eq::Zv}
\end{eqnarray}
where $C_A=N_c$, $C_F=(N_c^2-1)/(2N_c)$ and $T=1/2$ for a $SU(N_c)$ group and
the ellipses stand for non-fermionic and $\order{\alpha_s^4}$ terms.
Note that the strong coupling is defined in the effective theory with $n_l$
active quarks where $n_l+n_h$ is the total number of quark flavours.
In our case we have $n_h=1$, however, we keep $n_h$ in the formulae for
convenience.
Since there are poles starting from order $\alpha_s^2$, higher
order terms in $\epsilon$ are necessary for the decoupling relation of
$\alpha_s$. They have been computed in Ref.~\cite{Chetyrkin:1997un} and can be
found in explicit form in Eq.~(12) of Ref.~\cite{Grozin:2007fh}.
For the evaluation of $\Gamma_v$ to three-loop order
also one- and two-loop expressions for the 
strong-coupling and quark-mass counterterms are needed which have been
well-known for many years (see, e.g., Ref.~\cite{Marquard:2007uj}).

The approach which has been chosen for the evaluation of the Feynman diagrams
both at two-loop~\cite{Czarnecki:1997vz,Beneke:1997jm} and at three-loop
order~\cite{Marquard:2006qi} for the light-fermion contribution
is based on a partial fractioning of the integrands.
As a consequence the occurring integrals can be mapped to diagrams containing
less lines. Although they are in general simpler to evaluate they
contain denominators raised to higher powers which partially compensates this
advantage. 
On the other hand, the original vertex diagrams are more complicated to evaluate,
but we can use additional recurrence relations derived from the fact that
the momenta of the external quarks are the same.
Actually, we computed the three-loop $n_l$-contribution in
both ways and observed no significant difference in the overall performance.
Thus, in the automatic approach we refrain from performing the
partial fractioning and evaluate directly the vertex diagrams.

It is convenient to cast the perturbative expansion of the
matching coefficient in the form
\begin{eqnarray}
  c_v &=& 1 + \frac{\alpha_s^{(n_l)}(\mu)}{\pi} c_v^{(1)}
  + \left(\frac{\alpha_s^{(n_l)}(\mu)}{\pi}\right)^2 c_v^{(2)}
  + \left(\frac{\alpha_s^{(n_l)}(\mu)}{\pi}\right)^3 c_v^{(3)}
  + {\cal O}(\alpha_s^4)
  \,,
  \label{eq::cvdef}
\end{eqnarray}
where we further decompose $c_v^{(3)}$ according to the colour structures 
as
\begin{eqnarray}
  c_v^{(3)} &=& C_F T n_l\left(
  C_F\, c_{FFL} + C_A\, c_{FAL} + T n_h\, c_{FHL} + T n_l\, c_{FLL} 
  \right)
  \nonumber\\&&\mbox{}
  + C_F T n_h\left(
  C_F\, c_{FFH} + C_A\, c_{FAH} + T n_h\, c_{FHH}
  \right)
  \nonumber\\&&\mbox{}
  + \mbox{non-fermionic and singlet terms}
  \label{eq::cv3lnl}
  \,.
\end{eqnarray}
The one-~\cite{KalSar} and
two-loop~\cite{Czarnecki:1997vz,Beneke:1997jm,Kniehl:2006qw} terms have
been known for more than ten years. 
More recently also the three-loop corrections proportional to $n_l$ became
available~\cite{Marquard:2006qi}. The corresponding results read\footnote{Note
  that in Ref.~\cite{Marquard:2006qi} the result has been expressed in terms
  of the six-flavour coupling whereas here we use $\alpha_s^{(5)}$.
  This explains the difference in the logarithmic part of the coefficient $c_{FHL}$.}
\begin{eqnarray}
  c_v^{(1)}&=&-2C_F\,,
  \nonumber\\
  c_v^{(2)}&=&\left(-\frac{151}{72}
    +\frac{89}{144}\pi^2
    -\frac{5}{6}\pi^2\ln2-\frac{13}{4}\zeta(3)\right)C_AC_F
  \nonumber\\&&\mbox{}
  +\left(\frac{23}{8}-\frac{79}{36}\pi^2
    +\pi^2\ln2-\frac{1}{2}\zeta(3)\right)C_F^2
  +\left(\frac{22}{9}-\frac{2}{9}\pi^2\right)C_FTn_h
  \nonumber\\&&\mbox{}
  +\frac{11}{18}C_FTn_l
  -\frac{1}{2}\left[4\beta_0+\pi^2\left(\frac{1}{2}C_A +
      \frac{1}{3}C_F\right)\right]C_F\Lmu
  \,,
  \nonumber\\
  c_{FFL} &=& 46.7(1) + \left( -\frac{17}{12} + \frac{61}{36} \pi^2 -
    \frac{2}{3} \pi^2 \ln 2 + \frac{1}{3} \zeta(3) \right) \Lmu +
  \frac{1}{18} \pi^2 \Lmu^2\,, \nonumber\\
  c_{FAL} &=& 39.6(1) + \left( \frac{181}{54} - \frac{67}{432} \pi^2 +
    \frac{5}{9} \pi^2 \ln 2 + \frac{13}{6} \zeta(3) \right) \Lmu +
  \left( \frac{11}{9} + \frac{1}{12} \pi^2 \right) \Lmu^2\,,
  \nonumber\\
  c_{FHL} &=& -\frac{557}{162} + \frac{26}{81} \pi^2
  + \left( -\frac{44}{27} + \frac{4}{27} \pi^2 \right) \Lmu
  \,, \nonumber\\
  c_{FLL} &=& -\frac{163}{162} - \frac{4}{27} \pi^2
  - \frac{11}{27} \Lmu - \frac{2}{9} \Lmu^2\,,
  \label{eq::cv3}
\end{eqnarray}
with $\Lmu=\ln(\mu^2/m_Q^2)$ and $\beta_0 = ( 11/3\, C_A - 4/3\, Tn_l )/4$.
In this paper we consider the contributions proportional to $n_h$, i.e. new
results for $c_{FFH}$, $c_{FAH}$ and $c_{FHH}$ are presented.
We separate the logarithmic contributions and write
\begin{eqnarray}
  c_{FFH} &=& \tilde{c}_{FFH} - \frac{1}{20}\pi^2 \Lmu \,, 
  \nonumber\\
  c_{FAH} &=& \tilde{c}_{FAH} + \left( \frac{121}{27} - \frac{11}{27}\pi^2
  \right) \Lmu\,,
  \nonumber\\
  c_{FHH} &=& \tilde{c}_{FHH}\,,
  \label{eq::cv3nh}
\end{eqnarray}
where the $\Lmu$ term in $c_{FFH}$ arises from Eq.~(\ref{eq::Zv}) and
the one in $c_{FAH}$ originates from the running of $\alpha_s$.

Before considering the heavy-fermion contribution 
let us in a first step consider the two-loop and light-fermion contribution in
order to get some confidence in our approach.
In Tab.~\ref{tab::nl} 
the results of our approach are compared to the ones of
Refs.~\cite{Marquard:2006qi}. The uncertainties given in the middle column 
are obtained by adding the numerical uncertainties of the individual master
integrals in quadrature. We observe an impressive agreement for all
coefficients which is in particular true for the analytically known coefficients
$\tilde{c}_v^{(2)}$, $\tilde{c}_{\rm FHL}$ and $\tilde{c}_{\rm
  FLL}$. Furthermore, in the case of $\tilde{c}_{\rm FFL}$ and $\tilde{c}_{\rm
  FAL}$ a more precise result is obtained as compared to the approach of
Ref.~\cite{Marquard:2006qi} where the Mellin-Barnes method has been used for
the evaluation of the non-trivial master integrals.

\begin{table}
  \begin{center}
  \begin{tabular}{c|c|c}
                & this work & Ref.~\cite{Marquard:2006qi} \\
    \hline
$\tilde{c}_v^{(2)}$ & $-42.5138(2)$ & $-42.5140$ \\
$\tilde{c}_{\rm FFL}$ & $46.692(1)$ & $46.7(1)$ \\
$\tilde{c}_{\rm FAL}$ & $39.623(1)$ & $39.6(1)$ \\
$\tilde{c}_{\rm FHL}$ & $-0.27029(4)$ & $-0.27025$ \\
$\tilde{c}_{\rm FLL}$ & $-2.46833(3)$ & $-2.46834$ \\
$\tilde{c}_v^{(3)}|_{n_l}$ & $
n_l\left(120.660(3) - 0.8228 n_l\right) $ & $n_l\left(121. - 0.8228 n_l\right)$ \\
  \end{tabular}
  \caption{\label{tab::nl}Two-loop and light-fermion contribution to $c_v$.
    The results obtained with our numerical approach are compared to the ones
    of Ref.~\cite{Marquard:2006qi}. We have chosen the value $10^{-3}$
    for the {\tt FIESTA} parameter {\tt IfCut}.
    $n_h$ is set to one in the last row.}
  \end{center}
\end{table}

The new results for the $n_h$ contribution are shown in Tab.~\ref{tab::nh}. 
In the column ``numerical results'' we show the numbers as obtained from the
setup described in the previous section, i.e. there has been no manual
interaction in the calculation of three-loop diagrams.
On the other hand, if the master integrals which are known 
analytically\footnote{The numerical agreement between the analytically known
  integrals and the results from {\tt FIESTA} is about four to five digits
  in the coefficient of the highest $\epsilon$ expansion term which enters the
  finite contribution of $c_v$.
}
are used we obtain the numbers presented in the third column of
Tab.~\ref{tab::nh}.
One observes only marginal improvements. For this reason we use for the
following discussion the numerical results for the master integrals.

\begin{table}
  \begin{center}
  \begin{tabular}{c|c|c}
     & numerical & semi-analytical \\
     & result    & result \\
    \hline
$c_{\rm FFH}|_{\rm log}$ & $-0.496(2)$ & $-0.494(1)$ \\
$\tilde{c}_{\rm FFH}$ & $-0.841(3)$ & $-0.840(2)$ \\
$\tilde{c}_{\rm FAH}$ & $-0.10(2)$ & $-0.09(2)$ \\
$\tilde{c}_{\rm FHH}$ & $0.05126(1)$ & $0.05124$ \\
$\tilde{c}_{\rm FHL}$ & $-0.27029(4)$ & $-0.27025$ \\
$\tilde{c}_v^{(3)}|_{n_h}$ & $-0.93(4)- 0.09010(1) n_l$ & 
$-0.92(4) - 0.09008 n_l$ \\
  \end{tabular}
  \caption{\label{tab::nh}Three-loop heavy fermion contribution to $c_v$.
    For the results in the middle column we use only numerical results for the
    master integrals whereas in the right columns all available analytical
    information is employed. We have chosen the value $10^{-3}$ for the
    {\tt FIESTA} parameter {\tt IfCut}.
    $n_h$ is set to one in the last row.}
  \end{center}
\end{table}

We have performed several checks on the correctness of our result and on the
stability of the numerical calculations.
In the Feynman rule for the gluon propagator 
we allow for a general gauge parameter
and perform an expansion up to the linear term before 
the reduction to master integrals.
The cancellation of the gauge parameter in the final result, after including
counterterm contributions, serves as a welcome check for
the correctness of our result.
Furthermore, we have checked that both the spurious poles of order
$1/\epsilon^4$ and $1/\epsilon^5$ and the $1/\epsilon^3$ and $1/\epsilon^2$ poles
in our final expression cancel with an accuracy of about $10^{-4}$.

\begin{figure}[t]
  \leavevmode
  \epsfxsize=\textwidth
  \epsffile{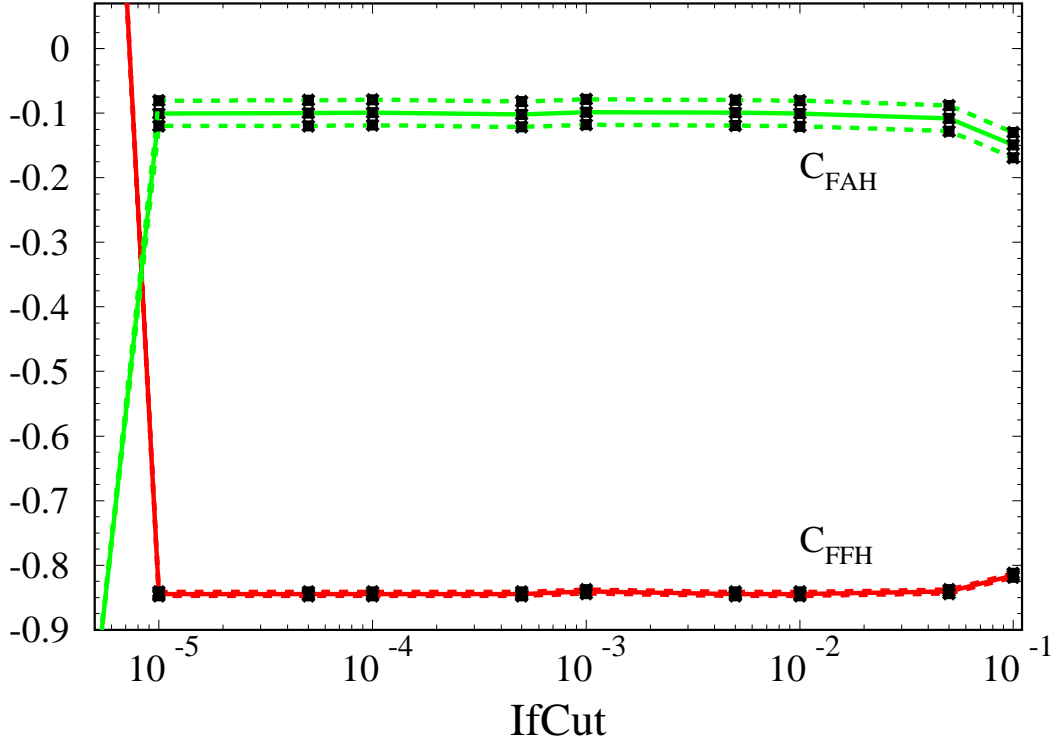}
\caption{\label{fig::nh}Coefficients $\tilde{c}_{FFH}$ and $\tilde{c}_{FAH}$ 
  as a function of {\tt IfCut} (choosing {\tt VarExpansionDegree=1}). 
  The band corresponds to the numerical
  uncertainty as provided by {\tt FIESTA}.}
\end{figure}

As a further test on the numerical stability of the evaluation of the master
integrals we vary the parameter {\tt IfCut}. In Fig.~\ref{fig::nh} the results
for $\tilde{c}_{FFH}$ and $\tilde{c}_{FAH}$ are shown for several values
between {\tt IfCut}$=5\cdot 10^{-5}$ and {\tt IfCut}$=0.1$ 
where for the guidance of the eye the 
data points are connected by straight lines. One observes a broad plateau with
only very minor variations. Larger deviations are obtained at the 
end points where either the value for {\tt IfCut} becomes too big or
numerical instabilities occur at the lower end of the integration region.
Note that the CPU time for the evaluation of the master integrals with {\tt
  FIESTA} varies from one to several days, depending on the setting for {\tt Vegas}.

As a further check on the numerical evaluation of the master integrals we have
used a different momentum assignment in the input for {\tt FIESTA}. As a
consequence different expressions are generated in intermediate steps leading
to different numerical integrations. The final results are in agreement with
the ones in Tab.~\ref{tab::nh} within a one sigma level.

A strong check on the numerical results for the master integrals is provided
by a change of the master integral basis. We replace the complicated integrals
by other 
integrals which we again evaluate with {\tt FIESTA}. The relations between the
old and new integrals can be extracted from the 
tables produced by {\tt crusher}.
By changing the basis two times (cf. Eqs.~(\ref{eq::MInum2})
and~(\ref{eq::MInum3})) we obtain the results given in
Tab.~\ref{tab::nh2}. 
We find good agreement with the results of
Tab.~\ref{tab::nh} for all coefficients.
Let us mention that in the standard basis spurious poles of at most fourth
order arise whereas both for basis~2 and~3 $1/\epsilon^5$ poles
are present in our result. Let us furthermore stress that the integrals for
the new bases are significantly more complicated due to the higher powers of
the propagators. This explains the slightly worse numerical
precision of the results in Tab.~\ref{tab::nh2}.

\begin{table}[t]
  \begin{center}
  \begin{tabular}{c|c|c}
     & basis 2 & basis 3 \\
     & Eq.~(\ref{eq::MInum2}) & Eq.~(\ref{eq::MInum3}) \\
    \hline
$c_{\rm FFH}|_{\rm log}$ & $-0.50(1)$ & $-0.496(8)$ \\
$\tilde{c}_{\rm FFH}$ & $-0.85(6)$ & $-0.86(2)$ \\
$\tilde{c}_{\rm FAH}$ & $-0.15(9)$ & $-0.13(4)$ \\
$\tilde{c}_{\rm FHH}$ & $0.0513(1)$ & $0.0513(1)$ \\
$\tilde{c}_{\rm FHL}$ & $-0.27028(3)$ & $-0.2703(2)$ \\
$\tilde{c}_v^{(3)}|_{n_h}$ & $-1.04(23) - 0.09009(1) n_l $ & 
    $-1.00(11) - 0.09009(5) n_l$ \\
  \end{tabular}
  \caption{\label{tab::nh2}Three-loop heavy fermion contribution to $c_v$.
    For the master integrals which are only available in numerical form
    the sets given in Eqs.~(\ref{eq::MInum2}) and~(\ref{eq::MInum3}) have been
    used. We have chosen the value $10^{-3}$ for the {\tt FIESTA}
    parameter {\tt IfCut}.
    $n_h$ is set to one in the last row.}
  \end{center}
\end{table}

Our final result for $c_v^{(3)}$ reads
\begin{eqnarray}
  \tilde{c}_{FFH} &=& -0.841(6)\,, \nonumber\\
  \tilde{c}_{FAH} &=& -0.10(4)\,, \nonumber\\
  \tilde{c}_{FHH} &=& -\frac{427}{162} + \frac{158}{2835}\pi^2 +
  \frac{16}{9}\zeta(3) \,\,\approx\,\, 0.05124\,,
  \label{eq::cv3nhtil}
\end{eqnarray}
where the values for $\tilde{c}_{FFH}$ and $\tilde{c}_{FAH}$ are taken
from Tab.~\ref{tab::nh}.
The uncertainties are conservatively estimated by doubling the error
from the numerical integration. In this way we account for effects
connected to the {\tt FIESTA} parameter {\tt IfCut}, to different momenta
assignments in the input of {\tt FIESTA}, 
and the change of the master integral basis.
The need for doubling the error can also be seen by
comparing the two values for $\tilde{c}_{FHH}$ in Tab.~\ref{tab::nh}.

Inserting the numerical values for the
colour factors we obtain for $\mu=m_Q$
\begin{eqnarray}
  c_v^{(3)} &\approx& -0.823\, n_l^2 +120.66(1)\, n_l - 0.93(8)
  + \mbox{non-fermionic and singlet terms}
  \,.
  \label{eq::cv3num}
\end{eqnarray} 
It turns out the numerical coefficient of the heavy-fermion contribution is
comparable with the $n_l^2$ part, however, significantly smaller
than the coefficient of the linear $n_l$ term.


\section{\label{sec::concl}Conclusions and outlook}

In this paper we discussed a completely automated approach for the calculation
of three-loop vertex corrections contributing to the matching coefficients of
the vector current, $c_v$. In particular we consider the Feynman diagrams
containing 
a closed heavy fermion loop which lead to significantly more complicated
integrals than the light-fermion contributions considered more than two years
ago~\cite{Marquard:2006qi}.
We have shown that numerically stable results are obtained even for the case
where all $\epsilon$ coefficients of all 24 master integrals are
evaluated numerically.
Furthermore, we were able to improve the precision of the light-fermion
contribution.
The method developed in this work will be crucial for the three-loop
non-fermionic and 
singlet contributions to $c_v$ which are still unknown.

Our automated approach depends crucially on the fact that {\tt FIESTA} is able
to evaluate the master integrals to a sufficiently high accuracy. To
make sure of this, we have checked that {\tt FIESTA} reproduces all
analytically known results within the given errors. In addition, we have
performed the calculation of the matching coefficient in three different
master integral bases and find consistent results. This makes us
confident that our result is correct within the given error bar.


\vspace*{1em}

\noindent
{\large\bf Acknowledgements}\\
We would like to thank A.~Czarnecki and A.A.~Penin for useful comments
and discussions and A.V.~Smirnov for continuous support on {\tt FIESTA}.
This work was supported by the DFG through SFB/TR~9.
The work of JHP and DS is supported by Science and Engineering Research
Canada and the Alberta Ingenuity Fund.
The Feynman diagrams were drawn with the help of
{\tt Axodraw}~\cite{Vermaseren:1994je} and {\tt
  JaxoDraw}~\cite{Binosi:2003yf}.


\section*{\label{sec::app}Appendix: master integrals}

In this appendix we list the 24 master integrals which we encounter in the
calculation for the $n_h$ contribution to $c_v$. We refrain from providing
explicit results since they can either be 
found in analytical form in the cited literature or can
be obtained in numerical form using {\tt FIESTA}~\cite{Smirnov:2008py}.

\begin{figure}
  \begin{center}
    \begin{tabular}{ccc}
      \includegraphics[width=9em]{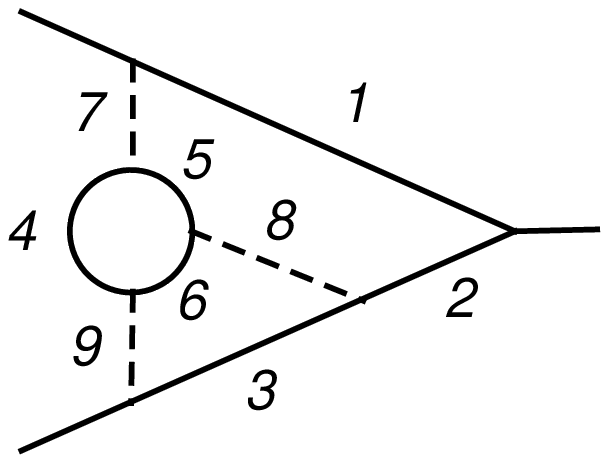} &
      \includegraphics[width=9em]{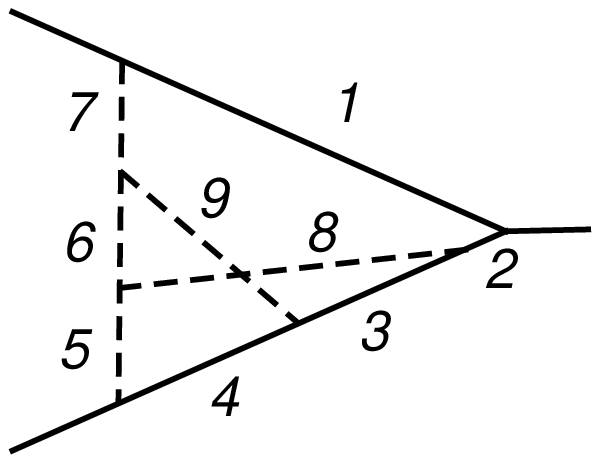} &
      \includegraphics[width=9em]{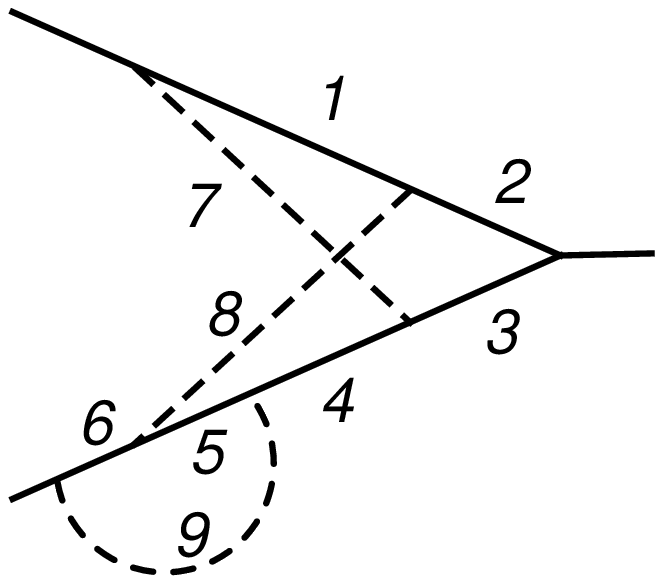}
      \\
      np2b2 & np12 & np33
    \end{tabular}
  \caption{\label{fig::MI}Generic vertex diagrams where
    the solid and dashed lines correspond to massive and massless propagators,
    respectively. The external momenta are set to $4m_Q^2$ (out-going on the
    right-hand side) and $m_Q^2$ (in-coming on the
    left-hand side), respectively.
    For $\VerttopOSlnp$ and $\VerttopOSlnpnp$ closed fermion contributions are
    generated by contracting massless lines.
  }
  \end{center}
\end{figure}

There are twelve master integrals which are known in analytical form to a
sufficiently high power in the $\epsilon$ expansion. They are given by 
\begin{eqnarray}&&
  \VerttopOSlnpb(0,1,0,1,1,0,0,1,1)\,,
  \VerttopOSlnpb(0,1,0,1,1,1,0,1,1)\,,
  \VerttopOSlnpb(0,1,1,0,1,1,0,0,0)\,,\nonumber\\&&
  \VerttopOSlnpb(0,1,1,0,1,1,1,0,1)\,,
  \VerttopOSlnpb(0,1,1,1,1,0,0,0,0)\,,
  \VerttopOSlnpb(0,1,1,1,1,0,0,1,1)\,,\nonumber\\&&
  \VerttopOSlnpb(0,1,1,1,1,1,0,0,0)\,,
  \VerttopOSlnp(0,1,1,1,0,0,0,0,0)\,,
  \VerttopOSlnp(0,1,1,0,1,0,0,0,1)\,,\nonumber\\&&
  \VerttopOSlnp(0,1,1,0,1,1,0,0,0)\,,
  \VerttopOSlnp(0,1,1,1,0,1,0,0,0)\,,
  \VerttopOSlnp(1,0,1,1,0,0,0,1,0)\,,
  \nonumber\\
  \label{eq::MIana}
\end{eqnarray}
where $\VerttopOSlnpb$ and $\VerttopOSlnp$ are given in graphical form
in Fig.~\ref{fig::MI}. 
The numbers next to the lines mark the corresponding
indices. The results for these integrals can be found in
Refs.~\cite{Laporta:1996mq,Melnikov:2000zc} (see
also~\cite{Marquard:2006qi}). 

The poles of the integral $\VerttopOSlnpb(1,0,1,0,1,1,0,0,0)$ can be found in 
Ref.~\cite{Groote:2004qq}; the finite part is computed with {\tt FIESTA}.

The following eleven integrals are only known numerically
with the help of {\tt FIESTA}
\begin{eqnarray}&&
  \VerttopOSlnpb(1,0,1,0,1,1,1,0,1)\,,
  \VerttopOSlnpb(1,0,1,0,1,2,0,0,0)\,,
  \VerttopOSlnpb(1,0,1,1,0,1,0,1,0)\,,\nonumber\\&&
  \VerttopOSlnpb(1,0,1,1,0,1,0,2,0)\,,
  \VerttopOSlnpb(1,0,1,1,0,1,1,1,0)\,,
  \VerttopOSlnpb(1,0,1,1,1,1,0,0,0)\,,\nonumber\\&&
  \VerttopOSlnpb(1,0,1,1,1,1,0,1,0)\,,
  \VerttopOSlnpb(1,0,1,1,1,1,0,2,0)\,,
  \VerttopOSlnpnp(0,1,1,1,1,1,1,0,0)\,,\nonumber\\&&
  \VerttopOSlnpnp(1,1,0,1,1,1,1,0,0)\,,
  \VerttopOSlnpnp(1,1,1,0,1,1,0,0,1)\,.
  \label{eq::MInum}
\end{eqnarray}

The change of the master integral basis discussed in
Section~\ref{sec::match} affects only the integrals in Eq.~(\ref{eq::MInum}),
$\VerttopOSlnpb(1,0,1,0,1,1,0,0,0)$ 
and the last five integrals in Eq.~(\ref{eq::MIana})
which are replaced by either (``basis~2'')
\begin{eqnarray}&&
\VerttopOSlnp(0, 2, 1, 0, 1, 0, 0, 0, 1)\,, 
\VerttopOSlnp(0, 2, 1, 0, 1, 1, 0, 0, 0)\,, 
\VerttopOSlnp(0, 2, 1, 1, 0, 1, 0, 0, 0)\,, 
\nonumber\\&&
\VerttopOSlnp(1, 0, 2, 1, 0, 0, 0, 1, 0)\,, 
\VerttopOSlnp(2, 0, 1, 1, 0, 0, 0, 1, 0)\,, 
\VerttopOSlnpb(1, 0, 1, 0, 1, 1, 1, 0, 2)\,, 
\nonumber\\&&
\VerttopOSlnpb(1, 0, 1, 0, 1, 2, 1, 0, 1)\,, 
\VerttopOSlnpb(1, 0, 1, 0, 1, 3, 0, 0, 0)\,, 
\VerttopOSlnpb(1, 0, 1, 1, 0, 1, 0, 3, 0)\,, 
\nonumber\\&&
\VerttopOSlnpb(1, 0, 1, 1, 0, 1, 1, 2, 0)\,, 
\VerttopOSlnpb(1, 0, 1, 1, 0, 1, 2, 1, 0)\,, 
\VerttopOSlnpb(1, 0, 1, 1, 1, 1, 0, 3, 0)\,, 
\nonumber\\&&
\VerttopOSlnpb(1, 0, 1, 1, 1, 2, 0, 0, 0)\,, 
\VerttopOSlnpb(1, 0, 1, 1, 1, 2, 0, 1, 0)\,, 
\VerttopOSlnpb(1, 0, 1, 1, 1, 2, 0, 2, 0)\,, 
\nonumber\\&&
\VerttopOSlnpnp(0, 2, 1, 1, 1, 1, 1, 0, 0)\,, 
\VerttopOSlnpnp(2, 1, 0, 1, 1, 1, 1, 0, 0)\,, 
\VerttopOSlnpnp(2, 1, 1, 0, 1, 1, 0, 0, 1)\,,
\nonumber\\
  \label{eq::MInum2}
\end{eqnarray}
or (``basis~3'')
\begin{eqnarray}&&
\VerttopOSlnp(0, 1, 1, 0, 1, 0, 0, 0, 2)\,, 
\VerttopOSlnp(0, 1, 1, 0, 1, 2, 0, 0, 0)\,, 
\VerttopOSlnp(0, 1, 1, 1, 0, 2, 0, 0, 0)\,, 
\nonumber\\&&
\VerttopOSlnp(1, 0, 1, 1, 0, 0, 0, 2, 0)\,, 
\VerttopOSlnp(1, 0, 2, 1, 0, 0, 0, 1, 0)\,, 
\VerttopOSlnpb(1, 0, 1, 0, 1, 1, 2, 0, 1)\,, 
\nonumber\\&&
\VerttopOSlnpb(1, 0, 1, 0, 2, 1, 0, 0, 0)\,, 
\VerttopOSlnpb(1, 0, 1, 0, 2, 1, 1, 0, 1)\,, 
\VerttopOSlnpb(1, 0, 1, 1, 0, 2, 0, 1, 0)\,, 
\nonumber\\&&
\VerttopOSlnpb(1, 0, 1, 1, 0, 2, 0, 2, 0)\,, 
\VerttopOSlnpb(1, 0, 1, 1, 0, 2, 1, 1, 0)\,, 
\VerttopOSlnpb(1, 0, 1, 1, 1, 2, 0, 1, 0)\,, 
\nonumber\\&&
\VerttopOSlnpb(1, 0, 1, 1, 2, 1, 0, 0, 0)\,, 
\VerttopOSlnpb(1, 0, 1, 1, 2, 1, 0, 1, 0)\,, 
\VerttopOSlnpb(1, 0, 1, 1, 2, 1, 0, 2, 0)\,, 
\nonumber\\&&
\VerttopOSlnpnp(0, 1, 1, 1, 1, 1, 2, 0, 0)\,, 
\VerttopOSlnpnp(1, 1, 0, 1, 1, 1, 2, 0, 0)\,, 
\VerttopOSlnpnp(1, 1, 1, 0, 1, 1, 0, 0, 2)\,.
\nonumber\\
\label{eq::MInum3}
\end{eqnarray}
Note that both basis~2 and~3 contain one more master integral than our
standard basis. This is because the latter in principle also contains the
integral $\VerttopOSlnpb(1,0,1,1,1,2,0,1,0)$, however, for our application the
corresponding coefficient is zero.



\end{document}